\documentclass[aps,prb,twocolumn,groupaddress,floatfix]{revtex4}
\pdfoutput=1
\usepackage{graphicx}
\usepackage{dcolumn}
\usepackage{bm}
\usepackage{amsfonts}
\usepackage{amsmath}
\usepackage{color}

\begin{document}

\title{Engineering artificial graphene in a two-dimensional electron gas}

\author{Marco Gibertini}
\author{Achintya Singha}
\author{Vittorio Pellegrini}
\author{Marco Polini}
\email{m.polini@sns.it}
\affiliation{NEST-CNR-INFM and Scuola Normale Superiore, I-56126 Pisa, Italy}
\author{Giovanni Vignale}
\affiliation{Department of Physics and Astronomy, University of Missouri, Columbia, Missouri 65211, USA}
\author{Aron Pinczuk}
\affiliation{Department of Physics and Department of Applied Physics
and Applied Mathematics, Columbia University New York, USA}
\author{Loren N. Pfeiffer}
\author{Ken W. West}
\affiliation{Bell Laboratories, Alcatel-Lucent Inc., Murray Hill, NJ
07974, USA}
\thanks{Present address: Department of Electrical Engineering,
Princeton University, Princeton, NJ, USA.}

\begin{abstract}
At low energy, electrons in doped graphene sheets behave like massless Dirac fermions with a Fermi velocity which does not depend on carrier density. Here we show that modulating a two-dimensional electron gas with a long-wavelength periodic potential with honeycomb symmetry can lead to the creation of isolated massless Dirac points with tunable Fermi velocity. We provide detailed theoretical estimates to realize such artificial graphene-like system and discuss an experimental realization in a modulation-doped GaAs quantum well. Ultra high-mobility electrons with linearly-dispersing bands might open new venues for the studies of Dirac-fermion physics in semiconductors.
\end{abstract}

\maketitle

Graphene is a one-atom-thick two-dimensional (2D) electron system composed of Carbon atoms
on a honeycomb lattice~\cite{reviews}. The lattice has two inequivalent sites in the unit cell that are analogous to the two spin orientations of a spin-$1/2$ particle. This observation opens the way to an elegant description of electrons in graphene as particles endowed with a {\it pseudospin} degree-of-freedom~\cite{reviews}.  At low energy, electrons in graphene are described by a 2D massless Dirac fermion (MDF) Hamiltonian, ${\cal H}_{\rm D} = v_{\rm F} {\bm \sigma} \cdot {\bm p}$, where $v_{\rm F}$ is the bare Fermi velocity, which does not depend on carrier density, ${\bm p}$ is the 2D  momentum measured from the corners of the Brillouin zone, and ${\bm \sigma}$ is the pseudospin operator constructed with two Pauli matrices $\{\sigma^i,i=x,y\}$, which act on the sublattice pseudospin degree-of-freedom. It follows that the energy eigenstates are {\it chiral}, {\it i.e.} for a given ${\bm p}$ have pseudospins oriented either parallel (conduction band) or antiparallel (valence band) to ${\bm p}$.  The Dirac-like wave equation and the chirality of its eigenstates have a number of very intriguing implications \cite{reviews}. Clearly it would be highly desirable to have other materials in which low-energy quasiparticles have Dirac-like spectrum and a pseudospin degree-of-freedom. One candidate is represented by HgTe/Hg(Cd)Te quantum wells
where MDFs are predicted to arise at a critical quantum-well thickness~\cite{konig}.  More recently, 
Park and Louie~\cite{park_nanolett_2009} have proposed that MDFs can arise in any 2D electron gas (2DEG) if appropriately nanopatterned.

Here we present an independent approach to the realization of  ``artificial graphene" in a nanopatterned 2DEG. We provide theoretical evidence for the occurrence of linearly-dispersing energy bands in an artificially engineered honeycomb lattice and we demonstrate a remarkable dependence of the Fermi velocity on the strength of the external potential in this system. We also define the conditions that the external periodic potential and the electron density must satisfy in order to achieve artificial MDFs. Finally we present the  photoluminescence (PL) of the 2DEG confined in a high-mobility modulation-doped GaAs/AlGaAs quantum well where a nanopatterning with honeycomb symmetry is achieved by dry etching. We believe that the development of patterned 2DEGs with tunable parameters will offer unprecedented opportunities to study fundamental interactions of MDFs in high-moblity semiconductor structures.

We start our analysis by considering a 2DEG consisting of electrons with band mass $m_{\rm b} = 0.067~m$ ($m$ is the bare electron mass in vacuum) confined in a thin quantum well created in a GaAs/AlGaAs semiconductor quantum well. The 2DEG is subjected to a periodic external potential $V_{\rm ext}({\bm r})$ with honeycomb structure. For the numerical calculations we have used a 2D muffin-tin potential which is zero everywhere but in disks of radius $r$, where it takes the constant value $V_0$. The center-to-center distance between the disks is $a$, and the lattice constant is $a_0 =\sqrt{3}~a$ (note that in Ref.~\onlinecite{park_nanolett_2009} they use a triangular rather than a honeycomb lattice).

Because the typical values of $a_0$ are much larger than the GaAs lattice constant, the external periodic potential can be viewed as a long-wavelength superlattice, which creates minibands. These are found by solving the secular equation~\cite{solid-state-book}
\begin{equation}
{\rm det}\left[\frac{\hbar^2}{2m_{\rm b}}({\bm k} + {\bm  G})^2 \delta_{{\bm G},{\bm G}'} +  V({\bm G}-{\bm G}')\right] = 0~.
\end{equation}
Here $V({\bm G})$ are the Fourier components of the external potential,
\begin{equation}
V_{\rm ext}({\bm r}) =  \sum_{{\bm G}} V({\bm G}) \exp{(i {\bm G} \cdot {\bm r})}~,
\end{equation}
${\bm G} = \ell {\bm g}_1 + p {\bm  g}_2$ are the reciprocal lattice vectors (RLVs) with $\ell$ and $p$ integers,
and ${\bm g}_1, {\bm g}_2$ are primitive RLVs, ${\bm g}_1 = 2\pi (1,\sqrt{3})/(3 a)$ and ${\bm g}_2= 2\pi (1,- \sqrt{3})/ (3a)$.

\begin{figure}
\centering
\tabcolsep=0cm
\begin{tabular}{cc}
\includegraphics[width=0.5\linewidth]{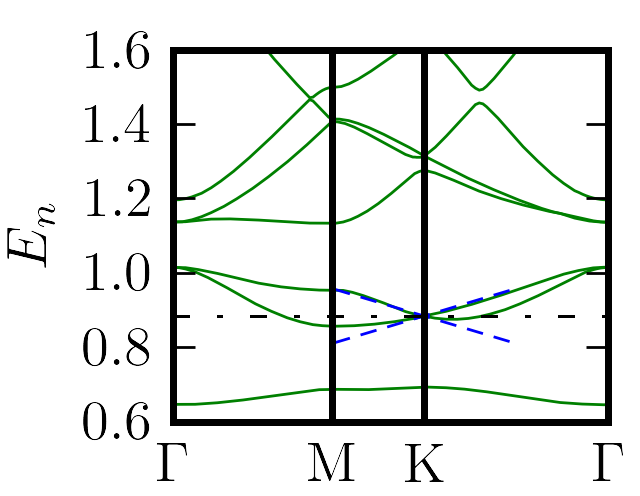} &
\includegraphics[width=0.5\linewidth]{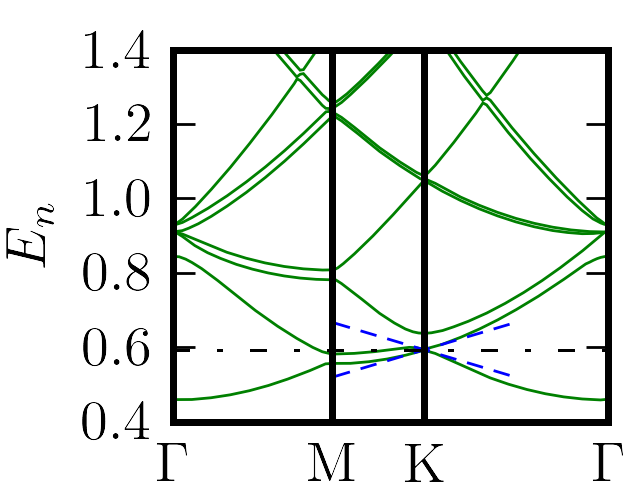} \\
\includegraphics[width=0.5\linewidth]{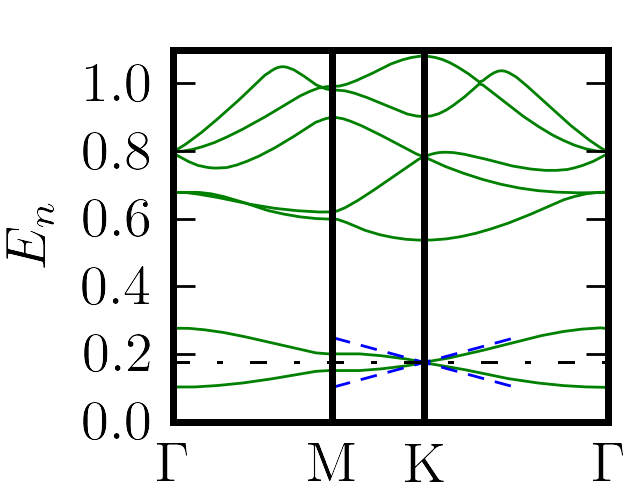} &
\includegraphics[width=0.5\linewidth]{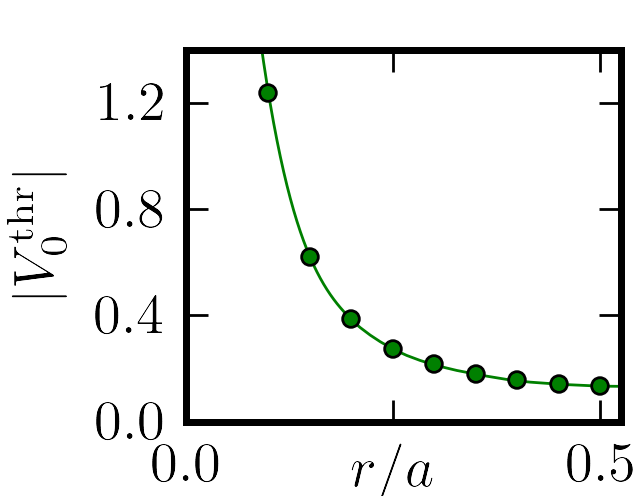}
\end{tabular}
\caption{(Color online) Top left panel: energy minibands (in ${\rm meV}$ and shifted upwards by $0.5~{\rm meV}$) for a repulsive potential with $V_0 = + 1.0~{\rm meV}$ [the points $\Gamma$, M, and K are defined in the inset to Fig.~\ref{fig:two}b)]. Dirac points are formed at the K point. The dashed lines are the approximate slopes of the bands near the Dirac points. Top right and bottom left: same as in the top left panel but for $V_0 = - 0.125~{\rm meV}$ and $V_0 = - 0.8~{\rm meV}$. Bottom right: the absolute value of the threshold potential $V^{\rm thr}_0$ (in ${\rm meV}$) as a function of $r/a$, for $a=150~{\rm nm}$.\label{fig:one}}
\end{figure}

In Fig.~\ref{fig:one} we plot the calculated energy minibands for a muffin-tin potential with $a=150~{\rm nm}$, $r= 52.5~{\rm nm}$, and for three different values of $V_0$: $V_0 = + 1.0~{\rm meV}$, $- 0.125~{\rm meV}$, and $- 0.8~{\rm meV}$.
As dictated by symmetry and group theory~\cite{slonczewski_1958}, these minibands are characterized by the existence of two-fold degenerate points at the corners of the Brillouin zone. It is easy to show that states with ${\bm k}$ close to these points are effectively described by a two-component MDF Hamiltonian ${\cal H}_{\rm D}$, with a Fermi velocity $v_{\rm F}$ that depends on $m_{\rm b}$, $a$, $V_0$, and on $r$ (see below). Comparing the different minibands shown in Fig.~\ref{fig:one}, we clearly see that in the case of a repulsive muffin-tin potential even when the Fermi level lies at the Dirac point (dash-dotted lines in Fig.~\ref{fig:one}), other type of states are present at the same energy. Ideally, similarly to what happens in graphene, one would like to be left only with isolated Dirac points at the Fermi level, {\it i.e.} with a gap in the bulk of the Brillouin zone. Within the muffin-tin model we have used, to create such gap we need an {\it attractive} potential, whose strength has to be stronger than a certain threshold $V^{\rm thr}_0$. The absolute value of $V^{\rm thr}_0$ is plotted in Fig.~\ref{fig:one} (bottom right panel)
as a function of the ratio $r/a$, for $a = 150~{\rm nm}$. For a geometry with $r/a \sim 0.35$ the threshold potential is $-0.18~{\rm meV}$.

Note that in the regime where isolated Dirac points exist, the band structure of the nanopatterned 2DEG at sufficiently low energies consists of manifolds of minibands separated by minigaps [see {\it e.g.} the bottom left panel in Fig.~\ref{fig:one}].

$V_0$, however, cannot be too large in absolute value. Indeed, when the local potential $V_0$ is attractive and too strong, it can lead to the formation of bound states. In this regime any small imperfection in the periodic structure, which is experimentally unavoidable, could lead to a dramatic change in the character of the states, yielding complete localization. Transport would occur mainly {\it via} variable-range hopping, a regime which we want to avoid. For a single disk we have estimated this threshold potential for the formation of bound states to be
\begin{equation}\label{eq:bound_states}
V^{\rm BS}_0 = - \beta^2~\frac{\hbar^2}{2 m_{\rm b} r^2} \sim -\frac{0.37}{(r/a)^2}~{\rm meV},
\end{equation}
where $\beta \sim 3.832$ is the first zero of the Bessel function $J_1(x)$. For $r/a \sim 0.35$ this localization threshold is roughly $3~{\rm meV}$, {\it i.e.} twenty times larger (in absolute value) than the threshold $|V^{\rm thr}_0|$ necessary to create a gap in the bulk of the Brillouin zone. Thus there is a precise window of values of $|V_0|$, which depends on $m_{\rm b}$ and on the geometrical parameters $a$ and $r$, which are suitable to create isolated Dirac points in the single-particle band structure of the 2DEG.

When a gap exists in the bulk of the Brillouin zone, the condition to have the Fermi level
exactly at the Dirac point is equivalent to the requirement of having just one band filled. The electron density $n_{\rm D}$ necessary to fill one band corresponds exactly to two electrons per unit cell, $n_{\rm D} = 4/( 3 \sqrt{3}~a^{2})$. For structures with $a \sim 150~{\rm nm}$ $n_{\rm D}$ is of the order of a few $10^{9}~{\rm cm}^{-2}$, which is a value that can be reached experimentally~\cite{hirjibehedin}. For electron densities $10^9~{\rm cm}^{-2} \leq n \leq 10^{10}~{\rm cm}^{-2}$, the appropriate $a$ changes from $277~{\rm nm}$ to $88~{\rm nm}$.

\begin{figure}
\centering
\includegraphics[width = 1.0\linewidth]{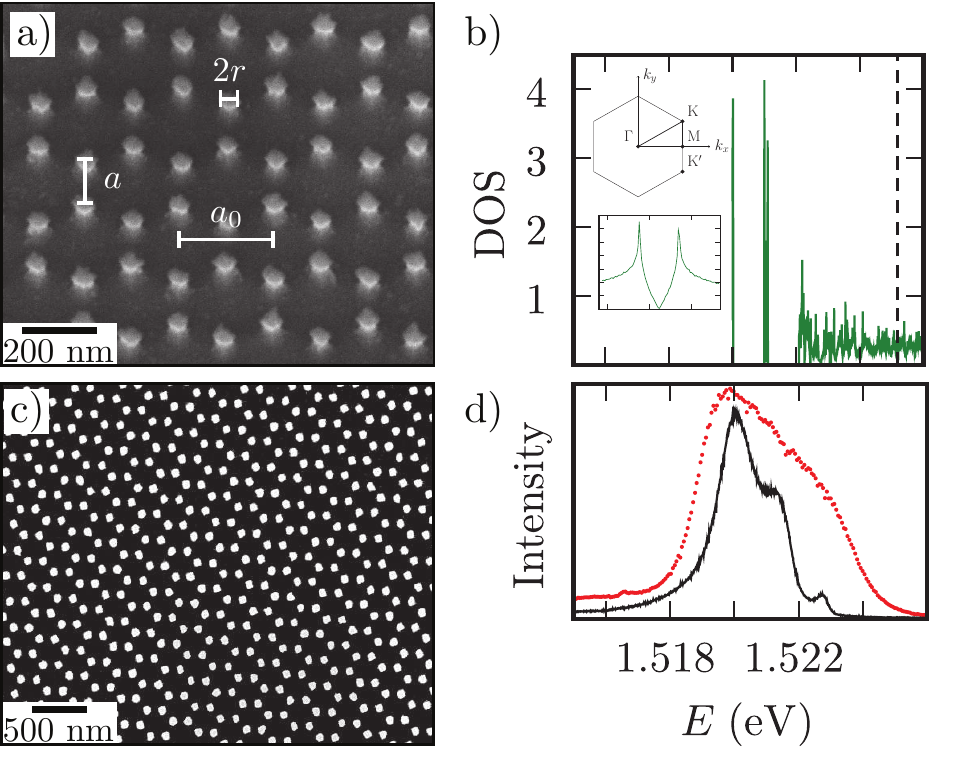}
\caption{(Color online) Panels a) and c) SEM images of the nanopatterned modulation-doped GaAs/AlGaAs sample. Panel b) Calculated density-of-states (DOS), in units of ${\rm eV}^{-1} \times {\rm nm}^{-2}$, as a function of energy for a muffin-tin potential with $V_0= -2.8~{\rm meV}$, $a=150~{\rm nm}$, and $r=52.5~{\rm nm}$. The first structure in the DOS is fixed at the energy of the main photoluminescence (PL) peak [see panel d)]. The vertical dashed line denotes the Fermi level associated with the nominal density $n_{\rm e} =1.1 \times 10^{11}~{\rm cm}^{-2}$. The insets show (i) the Brillouin zone corresponding to the external periodic potential represented in panel a) and (ii) a zoom of the DOS corresponding to the first structure in the main panel. The v-shaped DOS characteristic of MDFs is evident. Panel d) Low-temperature ($T=2~{\rm K}$) PL of the sample before (red dotted curve) and after (black solid curve) the processing.  \label{fig:two}}
\end{figure}
To explore the applicability of these ideas to real systems we realized the periodic external potential on a sample containing a 2DEG in a $25~{\rm nm}$ wide, one-side modulation-doped Al$_{0.1}$Ga$_{0.9}$As/GaAs quantum well. The 2DEG, positioned $170~{\rm nm}$ underneath the surface (the doping layer is at $110~{\rm nm}$), has measured low-temperature electron density $n_{\rm e} = 1.1 \times 10^{11}~{\rm cm}^{-2}$ and mobility of $2.7 \times 10^6~{\rm cm}^{2}/({\rm V s})$. The external modulation of the 2DEG is achieved following the procedure described in Ref.~\onlinecite{garcia} based on e-beam nanolithography and
inductive coupled plasma reactive ion shallow etching ($80~{\rm nm}$ below the surface).

Figs.~\ref{fig:two}a)  and~\ref{fig:two}c) show SEM images of the nanopatterned 2DEG. Other schemes with metallic gates can also be explored~\cite{superlattices}. The experimental values of the parameters are~\cite{metallurgic} $a \sim 150~{\rm nm}$ and $r \sim 50~{\rm nm}$, similar to those used for the calculations. $V_0$ is not known experimentally. However, one can engineer systems in which the potential felt by the electrons at the corners of the hexagonal cells is either repulsive ($V_0>0$) or, as in Fig.~\ref{fig:two}a) and c), attractive ($V_0<0$).

Since $n_{\rm e} > n_{\rm D}$, electrons occupy not only the bands with isolated Dirac points but also minibands at higher energy,
as shown in Fig.~\ref{fig:two}b), where the density-of-states (DOS) and the nominal Fermi level for $V_0 = -2.8~{\rm meV}$ are shown. Despite the large doping, the impact of the nanopatterning and the formation of minibands clearly manifest in the PL spectrum. Fig.~\ref{fig:two}d) shows representative PL spectra at $2~{\rm K}$ both of the unprocessed (red dotted curve) and processed (black solid curve) samples. In the unprocessed 2DEG case, the PL shape is determined by the density-of-states of the free electrons and equilibrium occupation factors of the 2DEG and photoexcited holes~\cite{pin} leading to an estimated electron density in agreement with the transport results. The processed sample PL, on the contrary, displays a remarkable change with the appearance of sharp structures on the high energy side and an overall reduction of its linewidth, which are remarkably consistent with the modification of the conventional constant-in-energy DOS as
shown in Fig.~\ref{fig:two}b), provided that $V_0$ is chosen appropriately~\cite{note} (at $V_0 = -2.8~{\rm meV}$ in this case). The overall smaller linewidth of the PL suggests a reduction of the average electron density due to the impact of the etching process.

We would like now to comment on the magnitude and tunability of the Fermi velocity of these artificially-induced MDFs. Using first-order perturbation theory it is possible to calculate analytically the slope of the bands close to the K point: in agreement with Ref.~\onlinecite{park_nanolett_2009} we find $v^{({\rm nf})}_{\rm F} = 2\pi \hbar/(3\sqrt{3} m_{\rm b} a)$, which is ({\it to this order of perturbation theory}) independent of the strength and the sign of the potential $V_0$ and is exactly one-half the velocity of a free electron of wave-vector K in the absence of the modulation. This ``nearly free" result applies only for $V_0 \to 0$. In passing, we note that using the value of the bare electron mass in vacuum $m_{\rm b} = m$, and $a = 1.42$~\AA~(which is the Carbon-Carbon distance in graphene) one gets $v^{({\rm nf})}_{\rm F} \sim 0.98 \times 10^{6}~{\rm m/s}$, which is surprisingly close to the Fermi velocity of electrons in graphene. Using instead the value of the band mass in GaAs, $m_{\rm b} \sim 0.067~m$, and $a = 150~{\rm nm}$, we find $v^{({\rm nf})}_{\rm F} \sim 1.4 \times 10^{5}~{\rm m/s}$, roughly an order of magnitude smaller than the value of the Fermi velocity in graphene.
\begin{figure}
\centering
\tabcolsep=0cm
\begin{tabular}{cc}
\includegraphics[width=0.5\linewidth]{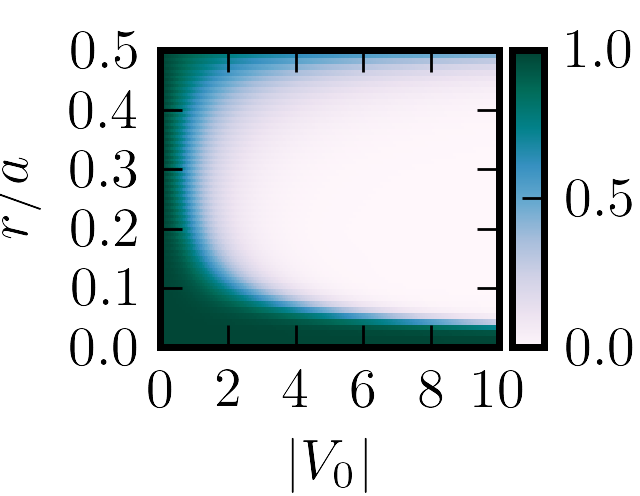} &
\includegraphics[width=0.5\linewidth]{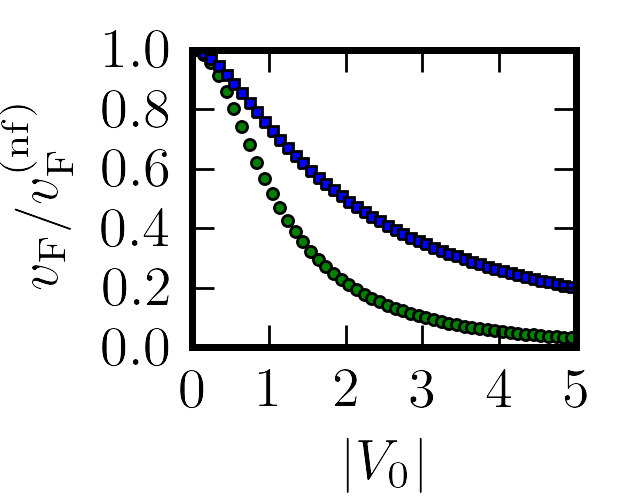}
\end{tabular}
\caption{(Color online) Left panel: color plot of the effective Fermi velocity $v_{\rm F}$ [in units of the nearly-free Fermi velocity $v^{({\rm nf})}_{\rm F}$] as a function of $r/a$ and $|V_0|$ (in ${\rm meV}$), for $a=150~{\rm nm}$. Right: $v_{\rm F}/v^{({\rm nf})}_{\rm F}$ as a function of $|V_0|$ for $r = 0.35$ (circles) and $r/a = 0.45$ (squares), again for $a=150~{\rm nm}$.\label{fig:three}}
\end{figure}
When $V_0$ is beyond the regime of applicability of first-order perturbation theory, the general formula for the Fermi velocity [in units of $v^{({\rm nf})}_{\rm F}$] is
\begin{equation}\label{vFermi}
\frac{v_{\rm F}}{v^{({\rm nf})}_{\rm F}} = 1 +  \frac{3\sqrt{3}~a}{2\pi} \sum_{\bm G}|w_0({\bm G})|^2 G_y~,
\end{equation}
where $w_0({\bm G})$ are the Fourier components of the periodic part of $w = 2^{-1/2}(\Psi_\uparrow - \Psi_\downarrow)$. Here $\{\Psi_\sigma({\bm r}), \sigma = \uparrow, \downarrow\}$ are the two degenerate eigenfunctions of the Hamiltonian ${\cal H} = {\bm p}^2/(2m_{\rm b}) + V_{\rm ext}({\bm r})$ at the K point, chosen to be a basis of the 2D representation of the little group presented in Ref.~\onlinecite{slonczewski_1958}. We have calculated Eq.~(\ref{vFermi}) numerically and the results are summarized in Fig.~\ref{fig:three}. We see that a finite value of $V_0$ (this plot concentrates only on $V_0 <0$), away from the regime of applicability of perturbation theory, tends to reduce the Fermi velocity with respect to $v^{({\rm nf})}_{\rm F}$. This effect is much stronger here than in Ref.~\onlinecite{park_nanolett_2009}: we attribute the difference to the different lattice structure (honeycomb vs triangular) and to the much smaller range of values of $V_0$ considered in Ref.~\onlinecite{park_nanolett_2009}.

Artificially-induced MDFs in 2DEGs confined in high-mobility semiconductor heterostructures could offer several advantages over graphene. One is clearly related to the very high purity of the former systems. Even though the dominant scattering mechanisms in graphene are not yet fully understood, it seems that charged impurities trapped close to (or on) the graphene plane play a very important role in limiting graphene's mobility~\cite{nomura_prl_2006,hwang_adam_prl_2007}, partly obscuring intrinsic properties of MDFs. The massless Dirac fluid at low densities is indeed a highly inhomogeneous system~\cite{jens_natphys_2008,yuanbo_condmat_2009,polini_prb_2008,rossi_prl_2008}. Although attempts to achieve high-mobilities in graphene systems~\cite{bolotin_ssc_2008,du_nature_nanotech_2008,orlita_prl_2008,graphene_on_graphite} seem to offer promising prospectives, the possibility of creating artificially MDFs in high-quality 2DEGs with mobilities that can exceed $10^7~{\rm cm}^2/({\rm V} {\rm s})$ even at low densities would represent a very exciting alternative route~\cite{nophonons}. The realization of artificial graphene thus would pave the way for the experimental observation of several predictions made for MDFs, such us a universal minimum conductivity~\cite{reviews} $\sigma_{\rm min} = 4 e^2/ (\pi h)$ and unusual electron-electron interaction physics~\cite{barlas_prl_2007,hwang_prl_2007,polini_ssc_2007,polini_arxiv_2009}.

Cyclotron-resonance in artificial graphene should exhibit a $\sqrt{B}$ dependence and quite large gaps. Indeed, the MDF cyclotron frequency is~\cite{reviews} $\omega^{\rm MDF}_{\rm c} = \sqrt{2} v_{\rm F}/\ell_B$, where $\ell_B= \sqrt{\hbar c/(eB)} \sim 257$~\AA$/\sqrt{B/{\rm Tesla}}$ is the magnetic length, while in a standard 2DEG it is instead $\omega^{\rm 2D~EG}_{\rm c} = \hbar/(m_{\rm b} \ell^2_B)$. In graphene $v_{\rm F} \sim 10^6~{\rm m/s}$ and thus, at a field $B = 10~{\rm T}$, $\omega^{\rm MDF}_{\rm c} \sim 1328~{\rm K}$. In GaAs, at the same field, $\omega^{\rm 2D~EG} \sim 63~{\rm K}$. In artificial graphene created in GaAs, as we have seen, the Fermi velocity is tunable: if we use its nearly-free value, $v_{\rm F} \sim 1.4 \times 10^5~{\rm m/s}$, we get $\omega^{\rm MDF}_{\rm c}  \sim 186~{\rm K}$ at $10~{\rm T}$, which is roughly one order of magnitude smaller than in graphene but still one order of magnitude larger than in a standard 2DEG. More importantly, since the fractional quantum Hall effect is easily observed in 2DEGs, artificial graphene could also be a very useful playground to understand electron-electron interaction effects in MDF systems in the presence of high magnetic fields. The large area of the superlattice unit cell offers also the possibility to achieve commensurability~\cite{hofstadter_prb_1976} between an external magnetic flux and the quantum of flux at quite small values of the external magnetic field.

The realization of MDFs in conventional semiconductors opens the interesting scenario related to the impact of spin-orbit coupling particularly if one uses InAs-based materials where spin-orbit coupling and effective $g$-factors are large. Finally we would like to mention that if the nanopatterning technique discussed above is carried out on areas with extension much smaller than that illustrated in Fig.~\ref{fig:two}, one can in principle create artificial-graphene ribbons with {\it perfect} zigzag or armchair edges in which confinement and boundary conditions could play a very important role.

In summary, we have shown that under suitable conditions systems of massless Dirac fermions can be created in conventional 2DEGs confined in semiconductor quantum wells by creating a superlattice with honeycomb geometry by nanopatterning. The existence of artificially-induced masslessness could be demonstrated through the observation of the half-integer quantum Hall effect~\cite{reviews} or by studying the PL or neutral collective excitations in the low-density regime. Such artificial graphene structures embedded in semiconductors could open novel routes for studies of electron interactions in low-dimensional systems.

\noindent{\it Acknowledgements}---M.P. acknowledges partial financial support from the CNR-INFM ``Seed Projects". G.V. acknowledges support from NSF Grant No. DMR-0705460. A.P. is supported by the Nanoscale Science and Engineering Initiative of the National Science Foundation under NSF Award Number CHE-0117752 and CHE-0641523, and by the New York State Office of Science, Technology, and Academic Research (NYSTAR).


\begin{thebibliography}{77}
\bibitem{metallurgic}
   This value of $r$ is the ``geometric" value: the actual value seen by the electrons buried in the quantum well is smaller.
\bibitem{note}
   We expect that the inclusion of the hole states in the description of the PL will not change the result of this analysis since hole
   minibands will be very flat due to the much larger hole mass.
\bibitem{nophonons}
   Note that there are neither lattice vibrations nor ripples associated with the artificially-created long-wavelength superlattice.
\end{thebibliography}
\end{document}